\documentclass[12pt]{article}
\usepackage{amsmath}
\usepackage{amssymb}
\usepackage{graphicx}

\def \noi {\noindent}
\def \bsk{\bigskip}
\def \msk{\medskip}

\begin{document}

\vspace*{2cm}
\begin{center}
{\large\bf EMPLOYING COMBINATION PROCEDURES TO SHORT-TIME EOP PREDICTION}
\end{center}

\vspace*{1cm}
\centerline{Z. MALKIN}
\centerline{Central Astronomical Observatory at Pulkovo of RAS}
\centerline{Pulkovskoe Ch. 65, St. Petersburg 196140, Russia}
\centerline{e-mail: malkin@gao.spb.ru}

\vspace*{1.5cm}
\noi{\bf ABSTRACT.} 
A well known problem with Earth Orientation Parameters (EOP) prediction is that a prediction
strategy proved to be the best for some testing time span and prediction length may not remain the same for other time intervals.
In this paper, we consider possible strategies to combine
EOP predictions computed using different analysis techniques to obtain
a final prediction with the best accuracy corresponding to the smallest prediction error of input predictions.
It was found that this approach is most efficient for ultra-short-term EOP forecast.

\vspace{0.5em}
\noi {\bf Keywords:} Earth Orientation Parameters, prediction

\bsk\noi {\bf 1. INTRODUCTION}

\begin{figure}[!ht]
\centering
\includegraphics[clip,width=\hsize]{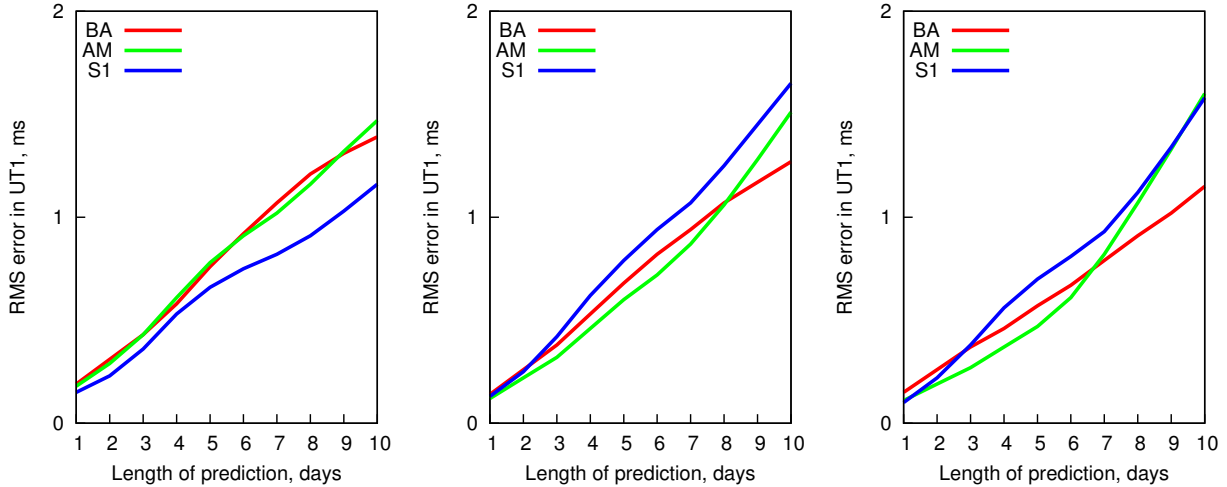}
\caption{Results of actual UT1 predictions for three years 2003, 2004 and 2005.
BA~--- Bulletin A prediction, AM~--- prediction of the same NEOS series made by the author,
S1~--- author's prediction of the SLR series computed at the Institute of Applied Astronomy.}
\label{fig:UT1-actual}
\end{figure}

\vspace{0.5em}\noi
Prediction of the Earth Orientation Parameters (EOP) is a practically very important and theoretically very interesting task, one of
the main fields of activity of operational EOP services.
Various methods are developed to compute a highly accurate EOP forecast.
However, a usual and well known problem is that different methods show different accuracy at different time intervals and prediction lengths.
A method which is the best for short-time prediction may not be such for long-time prediction, and vice versa.
On the other hand, a method which was proven to be the best for a testing period of time may not remain the same for the currently considered period.
As an example, results of actual Polar Motion (PM) and Universal Time (UT1) predictions are shown in Figure~\ref{fig:UT1-actual}.
In these plots, three predictions compared are made at U.S. Naval Observatory (BA~--- IERS Bulletin A), at Pulkovo Observatory
(AM~--- prediction of the IERS Bulletin A data), and S1~--- prediction of SLR series computed an the Institute of Applied
Astronomy\footnote{ftp://quasar.ipa.nw.ru/pub/EOS/IAA/erp\_rs.dat}.

\vspace{0.5em}
From this example, one can see that different predictions show the best results for different years and lengths of prediction.
This is true not only for different prediction methods, but also for variants of the same method.
All the methods practically used for EOP predictions use various configuration parameters, such as the length of reference interval,
the order of polynomial and number of harmonics for least squares adjustment, or the orders of autoregression and moving average for ARIMA.
A prediction procedure adjusted for some time period for computation of the best EOP forecast may not remain the best for the following time span.
Evidently, the main reason of this is that the Earth rotation is more complicated process than we are able to describe by our forecast models.

\vspace{0.5em}
So, the question is, whether we can develop a prediction strategy robust to irregular behavior of the Earth?
One of the possible ways to achieve this goal is making use of combination procedures applied to a set of predictions made with different methods.
Such an approach was tested during processing of the results of the EOP prediction comparison campaign
(Luzum et. al, 2007; Schuh et al., 2008, Kalarus et al., 2010), and showed a promising result.
Our purpose was to perform further investigation of possibility of improving the accuracy of EOP prediction using various combination procedures.
In this paper, we have performed several tests of computation of combined predictions of PM and UT1 based on a posteriori and actual predictions.

\bsk\noi {\bf 2. TEST DESCRIPTION AND RESULTS}

\vspace{0.5em}\noi
Three combination strategies proposed in (Malkin, 2009) were examined in this study using different input data:
\begin{description}
\item{C1} --- the average of input predictions as proposed and tested by Luzum et al. (2007), and Schuh et al. (2008).
\item{C2} --- the best previous prediction for given length. Normally, we compute EOP prediction on the day of the last observed epoch (``today'').
To compute $n$-day ahead prediction we examine the set of predictions made $n$ days ago and select one that predicts today EOP most accurately.
Then we compute our today $n$-day ahead prediction making use of the method used for computation of the best prediction made $n$ days ago.
\item{C3} --- the best yesterday prediction.  To compute this prediction we examine the set of predictions made yesterday and use for today prediction
the method which corresponds to the prediction made yesterday which predicts today EOP most accurately.
\end{description}

\vspace{0.5em}
Note that C2 and C3 predictions also use C1 results along with the input data.
Strictly speaking, C2 and C3 are not really combinations, just selections of the method that assures the best previous prediction is used to compute
the current one.
These three combination strategies were tested using two kinds of input data.

\vspace{0.5em}
The first set of input predictions consists of nine daily prediction series computed from the IERS C04 final EOP series starting from shifted
epochs from 1 October 2006 through 31 December 2007.
Predictions were computed making use of prediction strategy developed at the Institute of Applied Astronomy (Malkin \& Skurikhina, 1996).
The method is based on use of three prediction techniques:
\begin{enumerate}
\item  Least squares fitting for trend and several harmonics (LS);
\item  Autoregression (AR);
\item  Autoregressive Integrated Moving Average (ARIMA).
\end{enumerate}

\vspace{0.5em}
Final prediction is composed of several segments for different prediction length intervals computed using different methods and/or model parameters
and merged in one continuous series.
LS+AR combination was found to be the most effective for UT1 prediction, and LS+ARIMA combination proved to be the best for PM prediction.
Later, such an approach was developed by Niedzielski \& Kosek (2008)
The main model parameters that can be varied to adjust the prediction procedures are the following:
\begin{enumerate}
\item  base interval used for LS fitting;
\item  number and periods of harmonics;
\item  trend order;
\item  AR order;
\item  IMA (Integrated Moving Average) order.
\end{enumerate}

\begin{figure}[p]
\centering
\includegraphics[clip,width=0.95\hsize]{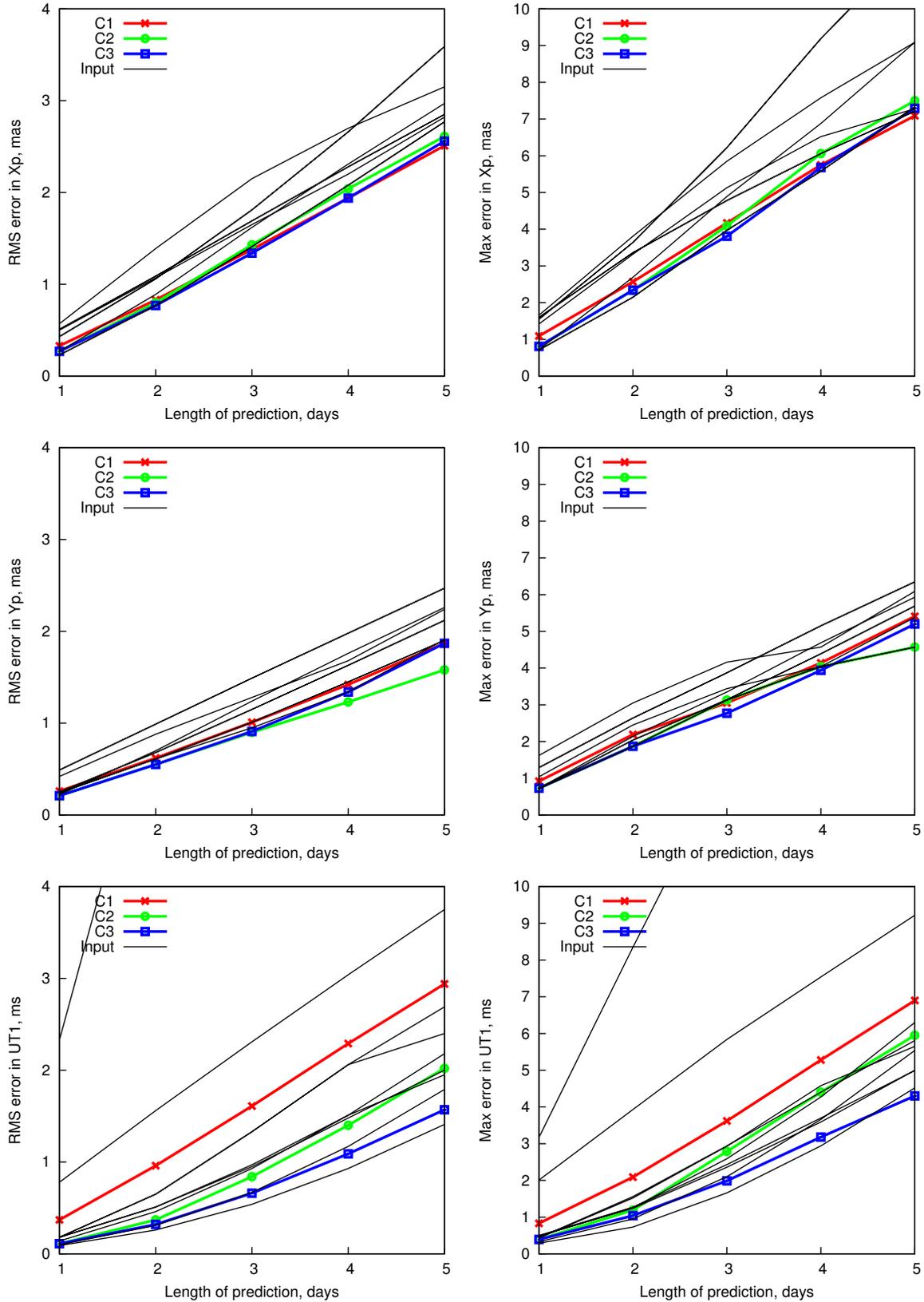}
\caption{The RMS and maximum errors of the input and combined predictions: a posteriori input predictions.}
\label{fig:test_simulated}
\end{figure}

\vspace{0.5em}
Thus, nine prediction series were computed using various sets of listed parameters aside from item 2 which remains the same for all the predictions.
Then three combined predictions C1, C2, and C3 were computed for the period from 1 January 2007 through 31 December 2007.
Figure~\ref{fig:test_simulated} shows the result of comparison of input and combined predictions with the IERS C04 EOP series.
Both the RMS and maximum differences between them are depicted in the plots.

\vspace{0.5em}
Hereafter we will consider ultra-short-term predictions with the length up to several days because the method discussed is mostly effective just
for such a kind of prediction.
Besides, ultra-short-term prediction is evidently the most required product.

\vspace{0.5em}
The results presented in Fig.~\ref{fig:test_simulated} clearly show that generally combined predictions provide high quality result,
more accurate than input series.
All three combined series show very similar accuracy for the PM predictions.
Maybe C3 solution has a marginal advantage.

\vspace{0.5em}
However, we have a much more complicated situation with the UT1 predictions.
The average (C1) prediction does not provide a satisfactory result, probably because of presence of bad predictions seen
in the central and top-left part of the plots related to UT1.
In this case, C3 (best yesterday prediction) is clearly the best for UT1 prediction.

\begin{figure}[p]
\centering
\includegraphics[clip,width=0.95\hsize]{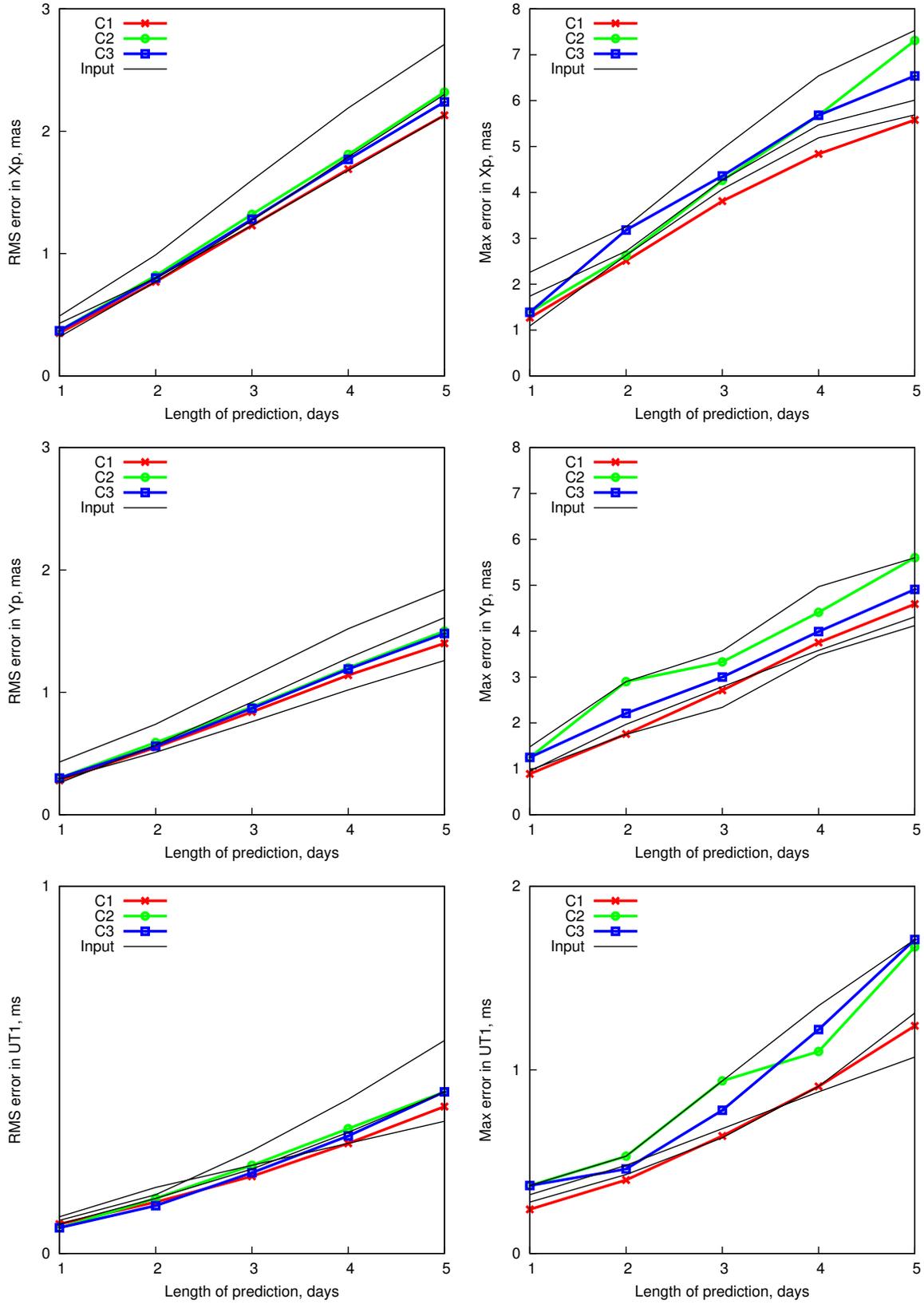}
\caption{The RMS and maximum errors of the input and combined predictions: actual input predictions.}
\label{fig:test_actual}
\end{figure}

\vspace{0.5em}
The second test was performed using the actual ERP predictions made at the U.S. Naval Observatory, Russian State EOP Service, and Pulkovo
Observatory in the period from January 2009 through June 2010.
Using these three input prediction series, three combined prediction series C1, C2 and C3 were computed as described above, and compared
with the IERS C04 EOP series.
Results of comparison, RMS and maximum differences of input and combined series with C04 are depicted in the plots in Fig.~\ref{fig:test_actual}.
Again, one can see an advantage of combined series with respect to the input ones.
In this case, however, the C1 solution (average prediction) has shown the best result, especially for the maximum prediction error.

\bsk\noi {\bf 3. CONCLUSIONS}

\vspace{0.5em}\noi
In this paper, we have studied three strategies of computation of combined predictions of Polar Motion and Universal Time.
Both a posteriori and actual input predictions were used for the experiment.
The obtained results have shown that a combined prediction can provide robust accurate operational EOP forecast.
Combined prediction is shown to be the most effective for ultra-short-term prediction for a length up to few days.
However, more detailed investigations are needed to make a justified choice between tested combination strategies.

\vspace{0.5em}
Two possible ways of practical use of combined prediction can be considered separately or jointly:
\begin{enumerate}
\item Combination of predictions computed at the same analysis center using different strategies.
\item Combination of predictions produced at different analysis centers.
\end{enumerate}
Moreover, combined predictions made using different combination strategies can be used as input for the second-stage combination.

\vspace{0.5em}
Indeed, to achieve the most accurate result using a combination procedure, input predictions should be computed using the best methods.
Using all the input prediction series without preliminary assessment of their quality can dilute the accuracy of final product.

\eject

\bsk\noi {\bf REFERENCES}

\leftskip=5mm
\parindent=-5mm

\msk
Kalarus M., Schuh H., Kosek W., Akyilmaz O., Bizouard~Ch., Gambis~D., Gross~R., Kumakshev~S., Kutterer~H., Mendes Cerveira~P.J., Pasynok~S., Zotov~L.,
Achievements of the Earth orientation parameters prediction comparison campaign. {\it J.~Geodesy}, Vol.~84, 587--596.

\msk
Luzum B., Wooden W., McCarthy D., Schuh H., Kosek~W., Kalarus~M., (2007).
Ensemble Prediction for Earth Orientation Parameters, {\it Geophysical Research Abstracts}, Vol.~9, EGU2007-A-04315.

\msk
Malkin Z. (2009). Improving short-term EOP prediction using combination procedures.
{\it in: Proc. Journ\'ees 2008: Astrometry, Geodynamics and Astronomical Reference Systems, Dresden, Germany, 22--24 Sep},
ed. by M.~Soffel, N.~Capitaine, 164--167.

\msk
Malkin Z., Skurikhina E. (1996). On Prediction of EOP, {\it Communications of the Institute of Applied Astronomy RAS}, No.~93.

\msk
Niedzielski T., Kosek W. (2008). Prediction of UT1-UTC, LOD and AAM $\chi_3$ by combination of least-squares and multivariate stochastic methods.
{\it J.~Geodesy}, Vol.~82, 83--92.

\msk
Schuh H., Kosek W., Kalarus M., Akyilmaz O., Gambis~D., Gross~R., Jovanovic~B.,
\mbox{Kumakshev~S.}, Kutterer~H., Mendes Cerveira~P.J., Pasynok~S., Zotov L. (2008).
Earth Orientation Parameters Prediction Comparison Campaign~--- first summary, {\it Geophysical Research Abstracts}, Vol.~10, EGU2008-A-07644.

\end{document}